\documentclass{llncs}
\usepackage[english]{babel}
\usepackage[utf8]{luainputenc}
\usepackage{amsmath,amssymb,url,tabularx,subfloat}
\usepackage{graphicx,tikz,rotating,yh,soul}
\usepackage{tikz-dependency}
\usepackage[final]{showkeys}
\definecolor{refkey}{rgb}{1,0,.75}
\definecolor{labelkey}{rgb}{1,0,.75}
\usepackage[all]{xy}
\xyoption{all}
\usepackage[colorinlistoftodos]{todonotes}
\usepackage[lined,ruled]{algorithm2e}
\newcolumntype{C}{>{\centering\arraybackslash}X}

\title{Text Classification Using Association Rules, Dependency Pruning and Hyperonymization}
\titlerunning{Text Classification Using Association Rules\ldots}

\author{Yannis Haralambous \and Philippe Lenca}
\institute{Institut Mines Telecom, Telecom Bretagne,
UMR~CNRS~6285 Lab-STICC\\
Technopôle Brest Iroise
CS 83818, 29238~Brest~Cedex~3, France\\
{\tt <surname>.<name>@telecom-bretagne.eu}}

\def\petitid#1{{\scalebox{.8}[1]{\small\texttt{#1}}}}

\date{\today}

\begin{document}
\maketitle
\begin{abstract}
We present new methods for pruning and enhancing itemsets for text classification via association rule mining. Pruning methods are based on dependency syntax and enhancing methods are based on replacing words by their hyperonyms of various orders. We discuss the impact of these methods, compared to pruning based on tfidf rank of words.
\end{abstract}

\section*{Introduction}


Automatic text classification is an important text mining task, due to the huge number of text documents that we have to manage daily. Text classification has a wide variety of applications such as Web document and email classification. Indeed, most of the Web news services daily provide a large number of articles making them impossible to be organized manually~\cite{lang_ICML_1995}. Automatic subject classification~\cite{cohen_AAAI_1996} and SPAM filtering~\cite{shami_etal_AAAI_1998} are two additional examples of the interest of automatic text classification.


Automatic text classification can be defined as below. Given a set of documents such that each document is labeled with a class value, learn a model that assigns a document with unknown class to one or more particular classes. This can also be done by assigning a probability value to each class or by ranking the classes.


A wide variety of classical machine learning techniques have been used for text classification. Indeed, texts may be represented by word frequencies vectors, and thus most of the quantitative data methods can be used directly on the notorious “bag-of-words” model (cf.~\cite{sebastiani_ACM-CS_2002,aggarwal_zhai_MTD_2012}).


Choosing a classifier is a multicriteria problem. In particular one has often to make a trade-off between accuracy and comprehensibility. In this paper, we are interested in both criteria with a deeper interest in comprehensibility. We are thus interested in rule-based approaches and especially in class association rules algorithms. Several studies have already successfully considered association rule-based approaches in text mining (e.g., \cite{ahonen_etal_TR_1997}, \cite{zaiane_antonie_ADC_2002}, \cite{cherfi_etal_PMAR_2009}, \cite{roche_etal_IIPWM_2004}). This framework is suitable for considering some statistical characteristics (e.g., high-dimensionality, sparsity\ldots) of the bag-of-words model where a document is represented as a set of words with their associated frequency in the document.



However a text is more than a set of words and their frequencies. Enhancing the bag-of-words approach with linguistic features has also attracted several works (e.g., \cite{jaillet_etal_IDA_2006,do_Master_2012,Kovacs:2008da,OrdonezSalinas:2010db},  \cite{Pado:2007bu,Lowe:2001wx,Curran:2002vm}, \cite{nivre_TR_2005,FerreriCancho:2004wd}).


We here propose a class association rules based approach enriched by linguistic knowledge. The paper is organized as follows: after introducing the techniques we are going to use (class association rules \S\,\ref{car-sec}, dependencies \S\,\ref{dar}, hyperonymization \S\,\ref{hc}) we describe our main algorithms (for training \S\,\ref{train-sec}, classifying \S\,\ref{classify-sec} and evaluating \S\,\ref{evaluate-sec}); follows the experimental section, where we give results obtained by tfidf pruning \S\,\ref{tfidf}, dependency-based pruning \S\,\ref{dep-sec} and hyperonymization \S\,\ref{hyper-sec}, and, finally, we end up by a conclusion and perspectives for future work \S\,\ref{con}.






\section{Proposed model for text classification}

Let a \emph{corpus} be a set $\mathbf{C}=\{D_1,\ldots,D_n\}$ of documents. Let $\mC$ be a set of classes. An \emph{annotated corpus} is a pair $(\mathbf{C},\mathrm{class})$ where $\mathrm{class}:\mathbf{C}\to\mC$ is a function that maps each document $D_i$ to a (predefined) class of $\mC$.

A document $D\in\mathbf{C}$ is a set of sentences $S$. The corpus $\mathbf{C}$ can be considered as a set of sentences $\mathbf{S}=\{S_1,\ldots,S_m\}$ if we go through the forgetful functor (which forgets the document to which the sentence belongs). Repeated sentences in the same document, or identical sentences in different documents are considered as distinct, i.e., there is a function $\iota\colon\mathbf{S}\to\mathbf{C}$ which restores the forgotten information. We extend the $\mathrm{class}$ function to $\mathbf{S}$ by $\mathrm{class}(S):=\mathrm{class}(\iota(S))$.

A sentence $S$ is a sequence of words $w$ (sometimes we will consider $S$ simply as a set, without changing the notation). Let $\mW=\bigcup_{S\in\mathbf{S}} \bigcup_{w\in S}\{w\}$ be the set of all words of~$\mathbf{C}$.

\medskip

\subsection{Class association rules and text classification}\label{car-sec}

Let $\mI$ be a set of objects called \emph{items} and $\mC$ a set of classes. A \emph{transaction} $T$ is a pair $(\{i_1,\ldots,i_n\},c)$, where $\{i_1,\ldots,i_n\}\subseteq\mI$ and $c\in\mC$. We denote by $\mT$ the set of transactions, by $\mathrm{items}(T)$ the set of items (or “itemset”) of $T$ and by $\mathrm{class}(T)$ the class of $T$.

Let $I$ be an itemset. The \emph{support} of $I$ is defined by
$$\textstyle\mathrm{supp}(I):=\frac{\#\{T\in\mT\mid I\subseteq \mathrm{items}(T)\}}{\#\mT}.$$
Let $\sigma\in[0,1]$ be a value called \emph{minimum support}. An itemset $I$ is called \emph{frequent} if its \emph{support} exceeds $\sigma$.

The \emph{confidence} of a transaction $t$ is defined as
$$\textstyle\mathrm{conf}(t):=\frac{\#\{T\in\mT\mid \mathrm{items}(t)\subseteq\mathrm{items}(T)\wedge\mathrm{class}(t)=\mathrm{class}(T)\}}{\#\{T\in\mT\mid \mathrm{items}(t)\subseteq\mathrm{items}(T)\}}.$$
Let $\kappa\in[0,1]$ be a value called \emph{minimum confidence}. A \emph{class association rule} (or “CAR”) $r=(\{i_1,\ldots,i_n\},c)$ \cite{CAR} is a transaction with frequent itemset and a \emph{confidence} exceeding $\kappa$.



To classify text with CARs, we consider words as being items, documents as being itemsets and pairs of documents and classes as being transactions. The advantage of this technique is that CARs can be easily understood and hence potentially improved by the user, especially if the classifier is tuned so that it produces humanly reasonable number of rules. Once the classifier is trained, to classify a new sentence we first find all CARs whose items are contained in the sentence, and then use an aggregation technique to choose a predominant class among those of the CARs we found.

An important issue of CARs is that the complexity is exponential with respect to the itemset size, and hence we need to keep it bounded in specific ranges, independently of the size of documents to classify. Using entire documents as transactions is computationally out of reach, therefore pruning techniques play an important rôle. Our approach consists in (a) restricting CARs to the sentence level, (b) prune sentences by using morphosyntactic information (cf. \S\,\ref{dar}) and modifying itemsets using semantic information (cf. \S\,\ref{hc}). 

\subsection{Itemset pruning using dependencies}\label{dar}

One can prune sentences either by using word frequencies (cf. \S\,\ref{tfidf}) or by using information obtained by morphosyntactic parsing (cf. \S\,\ref{dep-sec}). In this paper we introduce the latter approach, in the frame of dependency grammar.

\emph{Dependency grammar} \cite{tesniere1959,melcuk} is a syntactic theory, alternative to \emph{phrase-structure analysis} \cite{chomsky1957} which is traditionally taught in primary and secondary education. In phrase-structure syntax,  trees are built by grouping words into “phrases” (with the use of intermediate nodes NP, VP, etc.), so that the root of the tree represents the entire sentence and its leaves are the actual words. In dependency grammar, trees are built using solely words as nodes (without introducing any additional “abstract” nodes). A single word in every sentence becomes the \emph{root} (or \emph{head}) of the tree. An oriented edge between two words is a \emph{dependency} and is tagged by a representation of some (syntactic, morphological, semantic, prosodic, etc.) relation between the words. For example in the sentence “John gives Mary an apple,” the word “gives” is the head of the sentence and we have the following four dependencies:

\begin{center}\begin{dependency}
\begin{deptext}[row sep=-5pt,column sep=1cm]
John \& gives \& Mary \& an \& apple.\\
\end{deptext}
\deproot{2}{head}
\depedge{5}{2}{dobj}
\depedge{1}{2}{nsubj}
\depedge{3}{2}{iobj}
\depedge{4}{5}{det}
\end{dependency}
\end{center}

\noindent where tags nsubj, dobj, iobj, det denote “noun subject,” “direct object,” “indirect object” and “determinant.”


Let $S$ be a sentence and $\mD$ be the set of dependency tags: \{nsubj, ccomp, prep, dobj, \ldots\} A \emph{dependency} is a triple $(w_1,w_2,d)$ where $w_1,w_2\in S$ and $d\in\mD$. Let $\mathrm{Dep}(S)$ denote the set of dependencies of $S$ and $\mathrm{root}(S)$ the head of~$S$. Pruning will consist in defining a \emph{morphosyntactic constraint} $\phi$ i.e. a condition on dependencies (and POS tags) of words, the fulfillment of which is necessary for the word to be included in the itemset.

But before describing pruning algorithms and strategies, let us first present a second technique used for optimizing itemsets. This time we use semantic information. We propose to replace words by their hyperonyms, expecting that the frequencies of the latter in the itemsets will be higher than those of the former, and hence will improve the classification process.

\subsection{Hyperonymization}\label{hc}

The WordNet lexical database \cite{miller1995} contains sets of words sharing a common meaning, called \emph{synsets}, as well as semantic relations between synsets, which we will use to fulfill our goal. More specifically, we will use the relations of \emph{hyperonymy} and of \emph{hyperonymic instance}. The graph having synsets as nodes, and hyperonymic relations as edges, is connected and rooted: starting with an arbitrary synset, one can iterate these two relations until attaining a sink. {Note that} in the case of nouns it will invariably be the synset \petitid{00001740} \{entity\} while for verbs there are approx.\ 550 different verb sinks. 

Let $\mathbb W$ be the WordNet lexical database, $s\in \mathbb W$  a synset and $h:\mathbb W\to 2^{\mathbb W}$ the hyperonymic or hyperonymic instance relation. We define an \emph{hyperonymic chain} $\mathrm{CH}(s)$ as a sequence $(s_i)_{i\geq0}$ where $s_0=s$ and $s_i\in h(s_{i-1})$, for all $i\geq 1$. Hyperonymic chains are not unique since a given synset can have many hyperonyms. To replace a word by the most pertinent hyperonym, we have to identify the most significant hyperonymic chains of it.

The \emph{wn-similarity} project \cite{pedersen} has released synset frequency calculations  based on various corpora. Let $\mathrm{lf}(s)$ denote the logarithmic frequency of synset~$s$ in the BNC English language corpus~\cite{bnc} and let us arbitrarily add infinitesimally small values to the frequencies so that they become unique ($s\ne s'\Rightarrow\mathrm{lf}(s)\ne\mathrm{lf}(s')$). We use frequency as the criterion for selecting a single hyperonymic chain to represent a given synset, and hence define the \emph{most significant hyperonymic chain} $\mathrm{MSCH}(s)$ as the hyperonymic chain $(s_i)_{i\geq0}$ of $s$ such that $s_i=\argmax_{s\in h(s_{i-1})}\mathrm{lf}(s)\text{, for all $i\geq1$}.$ The chain $\mathrm{MSCH}(s)$ is unique thanks to the uniqueness of synset frequencies.

Our CARs are based on words, not synsets. Hence we need to extend MSCHs to words. Let $w$ be a lemmatized word. We denote by $\mathrm{Synsets}(w)\subset\mathbb W$ the set of synsets containing $w$. If the cardinal $\#(\mathrm{Synsets}(w))>1$ then we apply a standard disambiguation algorithm to find the most appropriate synset $s_w$ for $w$ in the given context. Then we take $(s_i)_i=\mathrm{MSCH}(s_w)$ and for each synset $s_i$ in this chain we define $h_i(w)=\mathrm{proj}_1(s_i)$ $(i>0)$, that is the projection of $s_i$ to its first element, which by WordNet convention is the most frequent word in the synset. The function vector $h_*:\mW\to\mW$ (with $h_0\equiv\mathrm{Id}$) is called \emph{hyperonymization}, and $h_i(w)$ is the \emph{$i$-th order hyperonym of $w$}.

\section{Operational implementations for document classification}

Our text classifier operates by first training the classifier on sentences and then classifying the documents by aggregating sentence classification. These two procedures are described in Sections~\ref{train-sec} and~\ref{classify-sec} respectively. Specific evaluation procedure is presented in Section~\ref{evaluate-sec}.


\subsection{Training}\label{train-sec}

\begin{algorithm}[tb]
\SetKwFunction{train}{Train}
\SetKwFunction{prune}{Prune}
\SetKwFunction{hyper}{Hyperonymize}
\SetKwFunction{lemma}{Lemmatize}
\SetKwFunction{apriori}{Apriori}
\SetKw{KwTo}{:=}
\SetKwProg{Fn}{}{:}{end}

\KwData{An annotated corpus $\mathbf{C}$, values of minimum support $\sigma$ and minimum confidence $\kappa$}
\KwResult{A set of CARs $\mR=(\{R_1,\ldots,R_N\},\mathrm{conf})$ where $\mathrm{items}(R_i)\subset \mW$, $\mathrm{class}(R_i)\in\mC$, and $\mathrm{conf}(R_i)$ is the confidence of rule $R_i$}
\Fn(){\train{$\mathbf{C}$, $\sigma$, $\kappa$}}{
$\mathbf{S}$ \KwTo $\mathrm{forgetful}(\mathbf{C})$;
$\mathbf{S}'$ \KwTo $\emptyset$\;
	\For{$S\in\mathbf{S}$}
	{
	$S'$ \KwTo \hyper(\prune(\lemma($S$)))\;
	$\mathrm{class}(S')$ \KwTo $\mathrm{class}(\iota(S))$\;
    $\mathbf{S}'$ \KwTo $\mathbf{S}'\cup \{S'\}$\;
	}
$\mR$ \KwTo \apriori($\mathbf{S}',\sigma,\kappa$)\;
}
\caption{Training}\label{train}
\end{algorithm}

The \texttt{Train} algorithm (cf. Alg.~\ref{train}) takes as input an annotated corpus $\mathbf{C}$ and values of minimum support $\sigma$ and minimum confidence $\kappa$. It returns a set of CARs together with their confidence values.

The first part of the algorithm consists in processing the corpus, to obtain efficient and reasonably sized transactions. Three functions are applied to every sentence:
\begin{enumerate}
\item \texttt{Lemmatize} is standard lemmatization: let $\mP$ be the set of POS tags of the \emph{TreeTagger} system \cite{treetagger} (for example, NP stands for “proper noun, singular”, VVD stands for “verb, past tense”, etc.), and let $\mW'$ be the set of lemmatized forms of $\mW$ (for example, “say” is the lemmatized form of “said”); then we define
$\lambda:\mW\to(\mW\cup\mW')\times\mP$,
which sends a word $w$ to the pair $(w',p)$ where $w'$ is the lemmatized form of $w$ (or $w$ itself, if the word is unknown to \emph{TreeTagger}) and $p$ is its POS tag.
\item \texttt{Prune} is a function which prunes the lemmatized sentence so that only a small number of (lemmatized) words (and POS tags) remains. Several sentence pruning strategies will be proposed and compared (cf. \S\,\ref{tfidf} and \ref{dep-sec}).
\item \texttt{Hyperonymize} is a function which takes the words in the pruned itemset and replaces them by the members of their most significant hyperonymic chains. Several strategies will also be proposed and compared (cf. \S\,\ref{hyper-sec}).
\end{enumerate}

The second part of Alg.~\ref{train} uses the  \emph{apriori} algorithm \cite{apriori} with the given values of minimum support and minimum confidence and output restrictions so as to generate only rules with item $c\in\mC$ in the consequent. It returns a set $\mR$ of CARs and their confidence.

It should be noted that this algorithm operates on individual sentences, hereby ignoring the document level.

\subsection{Classification}\label{classify-sec}

\begin{algorithm}[tb]
\SetKwFunction{classify}{Classify}
\SetKwFunction{prune}{Prune}
\SetKwFunction{hyper}{Hyperonymize}
\SetKwFunction{lemma}{Lemmatize}
\SetKwFunction{apriori}{Apriori}
\SetKwFunction{sort}{Sort}
\SetKw{KwTo}{:=}
\SetKwProg{Fn}{}{:}{end}

\KwData{A set of CARs $\mR$, a document $D_0$}
\KwResult{The predicted class $\mathrm{predclass}(D_0)$, variety $\beta$, dispersion $\Delta$}
\Fn(){\classify{$\mR$, $D_0$}}{
\For{$S\in D_0$}
{
\If{$\exists r\in \mR$ such that $\mathrm{items}(r)\subset S$}{
$R_{S}$ \KwTo $\displaystyle\argmax_{r\in\mR\wedge \mathrm{items}(r)\subset S}\mathrm{conf}(r)$\;}
}
$\mathrm{predclass}(D_0)$ \KwTo $\displaystyle\argmax_{c\in\mC}\sum_{\substack{S\in D_0\\\mathrm{class}(R_{S})=c}}\mathrm{conf}(R_{S})$\;
$\beta$ \KwTo $\#\{c\in\mC\mid (\mathrm{class}(R_{S})=c)\wedge (\mathrm{conf}(R_{S})>0)\}$\;
$\Delta$ \KwTo $\displaystyle\max_{c\in\mC}\sum_{\substack{S_i\in D_0\\\mathrm{class}(R_{S_i})=c}}\mathrm{conf}(R_{S_i})-\min_{c\in\mC}\sum_{\substack{S_i\in D_0\\\mathrm{class}(R_{S_i})=c}}\mathrm{conf}(R_{S_i})$\;
}
\caption{Classification}\label{classify}
\end{algorithm}

The \texttt{Classify} algorithm (cf. Alg.~\ref{classify}) uses the set of CARs produced by \texttt{Train} to predict the class of a new document $D_0$ and furthermore provides two values measuring the quality of this prediction: variety $\beta$ and dispersion $\Delta$.

The first part of the algorithm takes each sentence $S$ of the document $D_0$ and finds the most confident CAR that can be applied to it (i.e., such that the itemset of the rule is entirely contained in the itemset of the sentence). At this stage we have, for every sentence: a rule, its predicted class and its confidence.

Our basic unit of text in \texttt{Train} is sentence, therefore CARs generated by Alg.~\ref{train}  produce a class for each sentence of $D_0$. An aggregation procedure is thus needed in order to classify the document.
This is done by taking class by class the sum of confidence of rules and selecting the class with the highest sum.

Although this simple class-weighted sum decision strategy is reasonable, it is not perfect and may lead to wrong classification. This strategy will be \textit{optimally} sure and robust if (a) the number of classes is minimal, and (b) the values when summing up confidence of rules are sufficiently spread apart. The degree of fulfillment of these two conditions is given by the parameters \emph{variety}~$\beta$ (the number of classes for which we have rules), and \emph{dispersion} $\Delta$ (the gap between the most confident class and least confident one). These parameters will contribute to comparison among the different approaches we will investigate.

\subsection{Evaluation}\label{evaluate-sec}

\begin{algorithm}[tb]
\SetKwFunction{traincomparable}{TrainComparable}
\SetKwFunction{classify}{Classify}
\SetKwFunction{evaluate}{Evaluate}
\SetKwFunction{singleevaluate}{SingleEvaluate}
\SetKwFunction{prune}{Prune}
\SetKwFunction{findoptimal}{FindOptimal}
\SetKwFunction{hyper}{Hyperonymize}
\SetKwFunction{lemma}{Lemmatize}
\SetKwFunction{apriori}{Apriori}
\SetKwFunction{partition}{Partition}
\SetKwFunction{shuffle}{Shuffle}
\SetKwFunction{sort}{Sort}
\SetKw{KwTo}{:=}
\SetKwProg{Fn}{}{:}{end}

\KwData{An annotated corpus $\mathbf{C}$, initial values of minimal support $\sigma_0$ and confidence $\kappa_0$, standard number of rules $\rho_0$}
\KwResult{Values of average precision $\overline P$, recall $\overline R$, F-measure $\overline F$. Values of average number of rules $\rho$, variety ${\beta}$ and dispersion ${\Delta}$}
\Fn(){\singleevaluate{$\mathbf{C}$, $\sigma$, $\kappa$}}{
$(\mathbf{C}_1,\ldots,\mathbf{C}_{10})$ \KwTo \partition(\shuffle($\mathbf{C}$),10)\;
\tcc{tenfold cross validation}
\For{$I\in \{1,\ldots,10\}$}
{
($\mR_I$, $\beta_I$, $\Delta_I$) \KwTo \train($\mathbf{C}\setminus\mathbf{C}_1$, $\sigma$, $\kappa$)\;
\For{$D\in\mathbf{C}_I$}{
$\mathrm{predclass}(D)$ \KwTo \classify($\mR_I$, $D$)\;
}
\For{$c\in\mC$}{
$R_I(c)$ \KwTo $\frac{\#\{d\in\mathbf{C}_I\mid (\mathrm{predclass}(d)=c)\wedge(\mathrm{class}(d)=c)\}}{\#\{d\in\mathbf{C}_I\mid \mathrm{class}(d)=c\}}$\;
$P_I(c)$ \KwTo $\frac{\#\{d\in\mathbf{C}_I\mid (\mathrm{predclass}(d)=c)\wedge(\mathrm{class}(d)=c)\}}{\#\{d\in\mathbf{C}_I\mid \mathrm{predclass}(d)=c\}}$;
$F_I(c)$ \KwTo $\frac{2R_I(c)P_I(c)}{R_I(c)+P_I(c)}$\;
}
}
\For{$c\in\mC$}{
$(R(c),P(c),F(c))$ \KwTo $\frac{1}{10}\sum_{I=1}^{10}(R_I(c),P_I(c),F_I(c))$\;
}
$(\rho,\beta,\Delta)$ \KwTo $\frac{1}{10}\sum_{I=1}^{10}(\#\mR_I,\beta_I,\Delta_I)$\;
$(\overline R,\overline P,\overline F)$ \KwTo $\frac{1}{\#\mC}\sum_{c\in\mC}(R(c),P(c),F(c))$\;
}
\Fn(){\evaluate{$\mathbf{C}$, $\sigma_0$, $\kappa_0$, $\rho_0$}}{
$(\sigma,\kappa)$ \KwTo $\findoptimal(\mathbf{C},\sigma_0,\kappa_0,\rho_0)$\;
$(\overline R,\overline P,\overline F,\rho,\beta,\Delta)$ \KwTo $\singleevaluate(\mathbf{C},\sigma,\kappa)$\;
}
\caption{Evaluation}\label{evaluate}
\end{algorithm}

We evaluate the classifier (Alg.~\ref{evaluate}), by using 10-fold cross validation to obtain average values of recall, precision, F-measure, variety and dispersion. This is done by algorithm \texttt{SingleEvaluate}, once we specify values of minimal support and minimal confidence.

Comparing rule-based classification methods is problematic because one can always increase F-measure performance by increasing the number of rules, which results in overfitting them. To avoid this phenomenon and compare methods in a fair way, we fix a number of rules $\rho_0$ (we have chosen $\rho_0=1{,}000$ in order to produce a humanly reasonably readable set of rules) and find values of minimal support and confidence so that \mbox{F-measure} is maximal under this constraint.

Function \texttt{FindOptimal} will launch \texttt{SingleEvaluate} as many times as necessary on a dynamic grid of values $(\sigma,\kappa)$ (starting with initial values $(\sigma_0,\kappa_0)$), so that, at the end, the number of rules produced by \texttt{Train} is as close as possible to $\rho_0$ (we have used $\#\mR\in[\rho_0-2,\rho_0+2]$) and $\overline F$ is maximal.

\section{Experimental results on Reuters corpus}\label{corpus}

In this section, we investigate three methods: (a) pruning through a purely frequentist method, based on tfidf measure (\S\,\ref{tfidf}); (b) pruning using dependencies (\S\,\ref{dep-sec}); (c) pruning using dependencies followed by hyperonymic extension~(\S\,\ref{hyper-sec}). 

\subsection{Preliminaries}
In the Reuters \cite{reuters} corpus we have chosen the 7 most popular topics (GSPO~= sports, E12~= monetary/economic, GPOL~= domestic politics, GVIO~= war, civil war, GDIP~= international relations, GCRIM~= crime, law enforcement, GJOB~= labor issues) and extracted the 1,000 longest texts of each. 

The experimental document set is thus a corpus of 7,000 texts of length between 120 and 3,961 words (mean $398.84$, standard variation $169.05$). The texts have been analyzed with the Stanford Dependency Parser \cite{sdp} in collapsed mode with propagation of conjunct dependencies. 



\subsection{Tfidf-based corpus pruning}\label{tfidf}

Tfidf-based corpus pruning consists in using a \textit{classical} \texttt{Prune} function as defined in Alg.~\ref{tfidf-alg}. It will be our baseline for measuring performance of dependency- and hyperonymy-based methods.

\begin{algorithm}[tb]
\SetKwFunction{classify}{Classify}
\SetKwFunction{evaluate}{Evaluate}
\SetKwFunction{prune}{Prune}
\SetKwFunction{hyper}{Hyperonymize}
\SetKwFunction{lemma}{Lemmatize}
\SetKwFunction{apriori}{Apriori}
\SetKwFunction{partition}{Partition}
\SetKwFunction{shuffle}{Shuffle}
\SetKwFunction{tfidf}{Tfidf}
\SetKwFunction{sort}{Sort}
\SetKw{KwTo}{:=}
\SetKwProg{Fn}{}{:}{end}

\KwData{An annotated corpus (considered as a set of sentences) $\mathbf{S}$}
\KwResult{The pruned corpus $\mathbf{S}'$}
\Fn(){\prune{$\mathbf{S}$, $N$}}{
$\mathbf{S}'$ \KwTo $\emptyset$\;
\For{$S\in \mathbf{S}$}{
\For{$w\in S$}{
$\tfidf_S(w)$ \KwTo $\freq_S(w)\cdot\log\left(\frac{\#\{S\in\mathbf{C}\}}{\#\{S\in\mathbf{C}\mid w\in S\}}\right)$
}
$S'$ \KwTo $\emptyset$; $S_0$ \KwTo $S$\;
\For{$i\in\{1,\ldots,N\}$}{
$w'$ \KwTo $\displaystyle\argmax_{S_0}\tfidf_S(w)$\;
$S_0$ \KwTo $S_0\setminus \{w'\}$; $S'$ \KwTo $S'\cup\{w'\}$\;
}
$\mathbf{S}'$ \KwTo $\mathbf{S}'\cup\{S'\}$\;
}
}
\caption{Tfidf-based corpus pruning}\label{tfidf-alg}
\end{algorithm}

Note that this definition of the tfidf measure diverges from the legacy one by the fact that we consider not documents but sentences as basic text units. This is because we compare tfidf-generated CARs to those using syntactic information, and syntax is limited to the sentence level. Therefore, in order to obtain  a fair comparison, we have limited term frequency to the sentence level and our  “document frequency” is in fact a sentence frequency.

\begin{figure}[ht]
\centering
\resizebox{\textwidth}{!}{\includegraphics{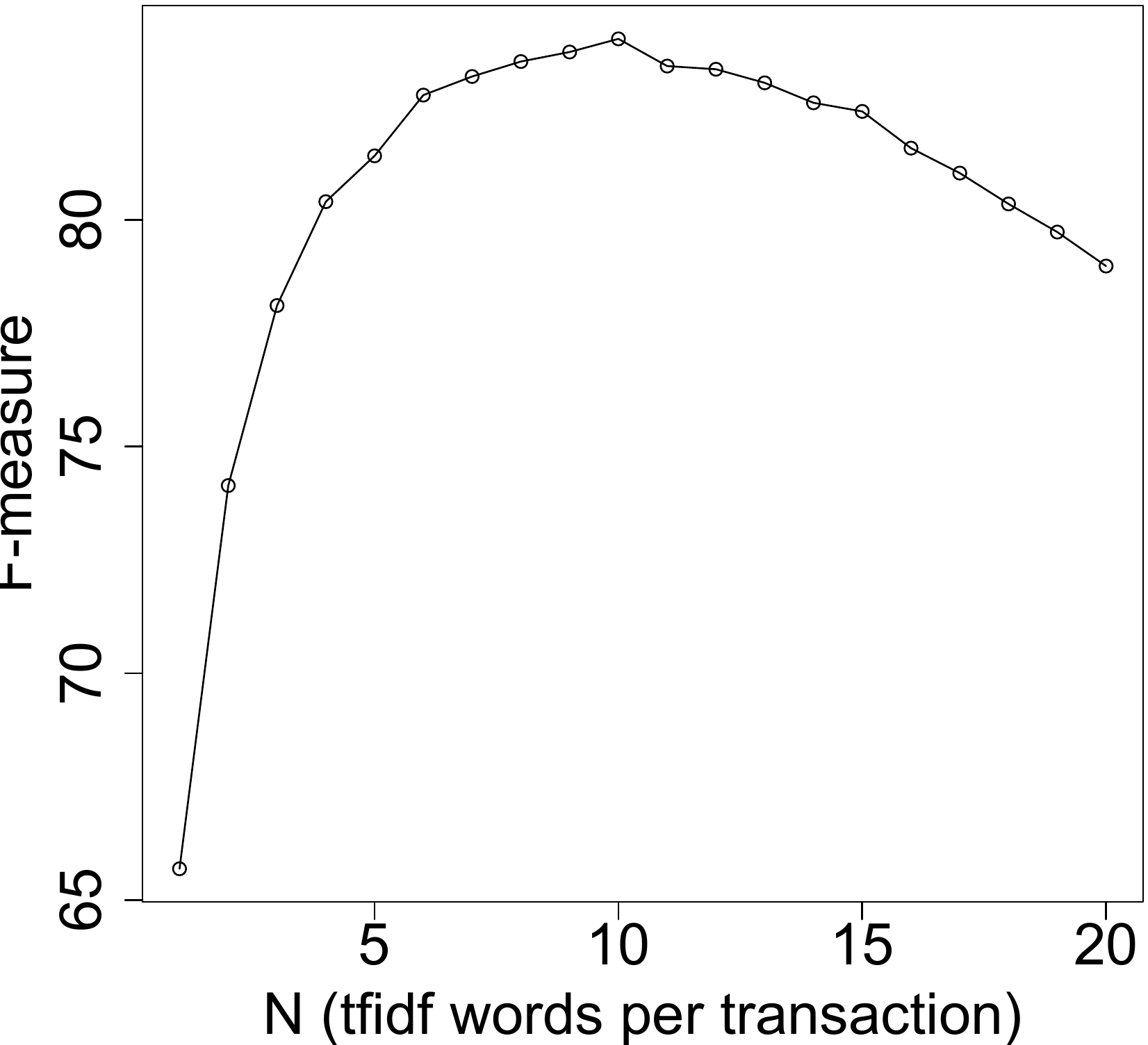}~\includegraphics{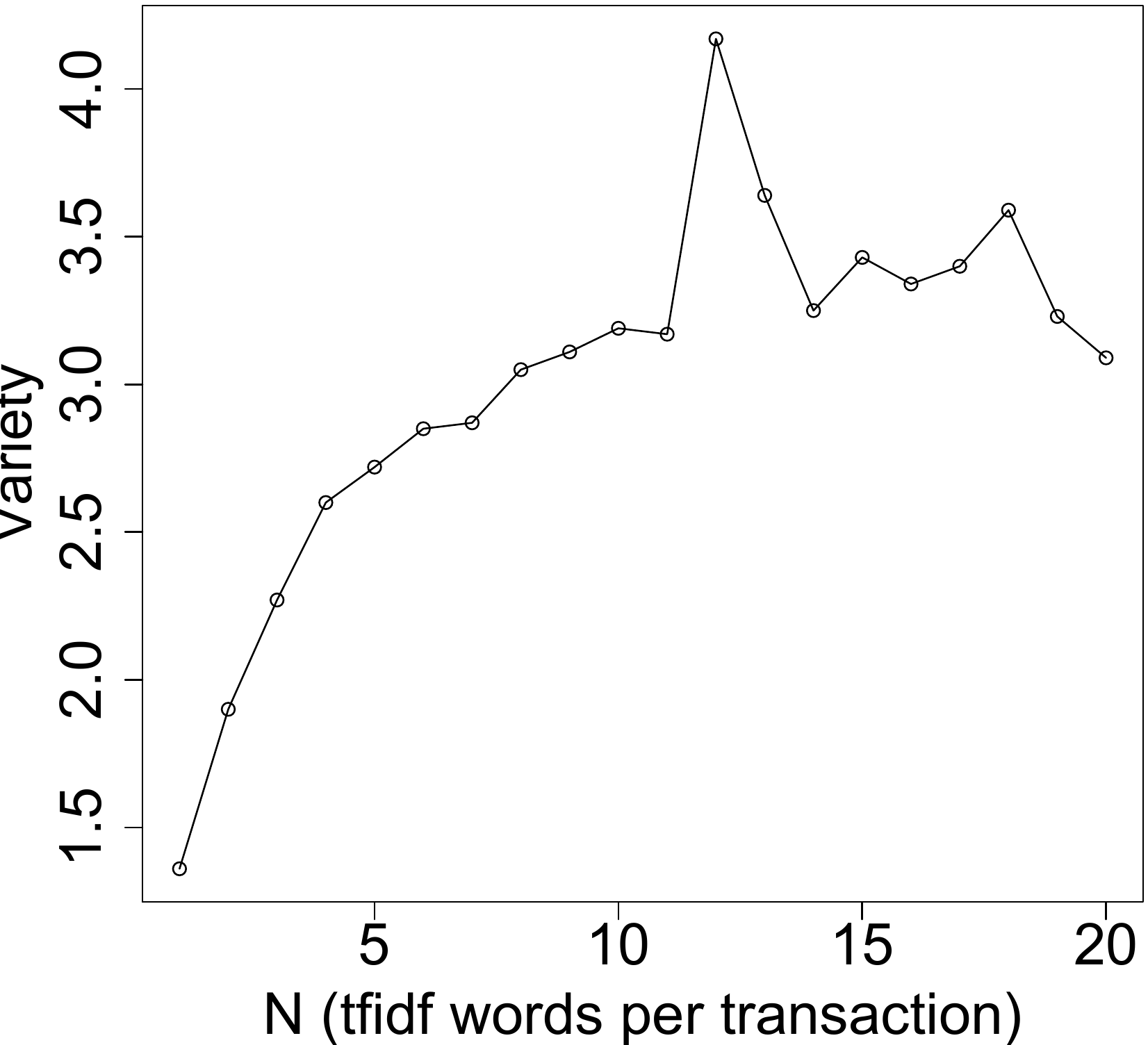}~\includegraphics{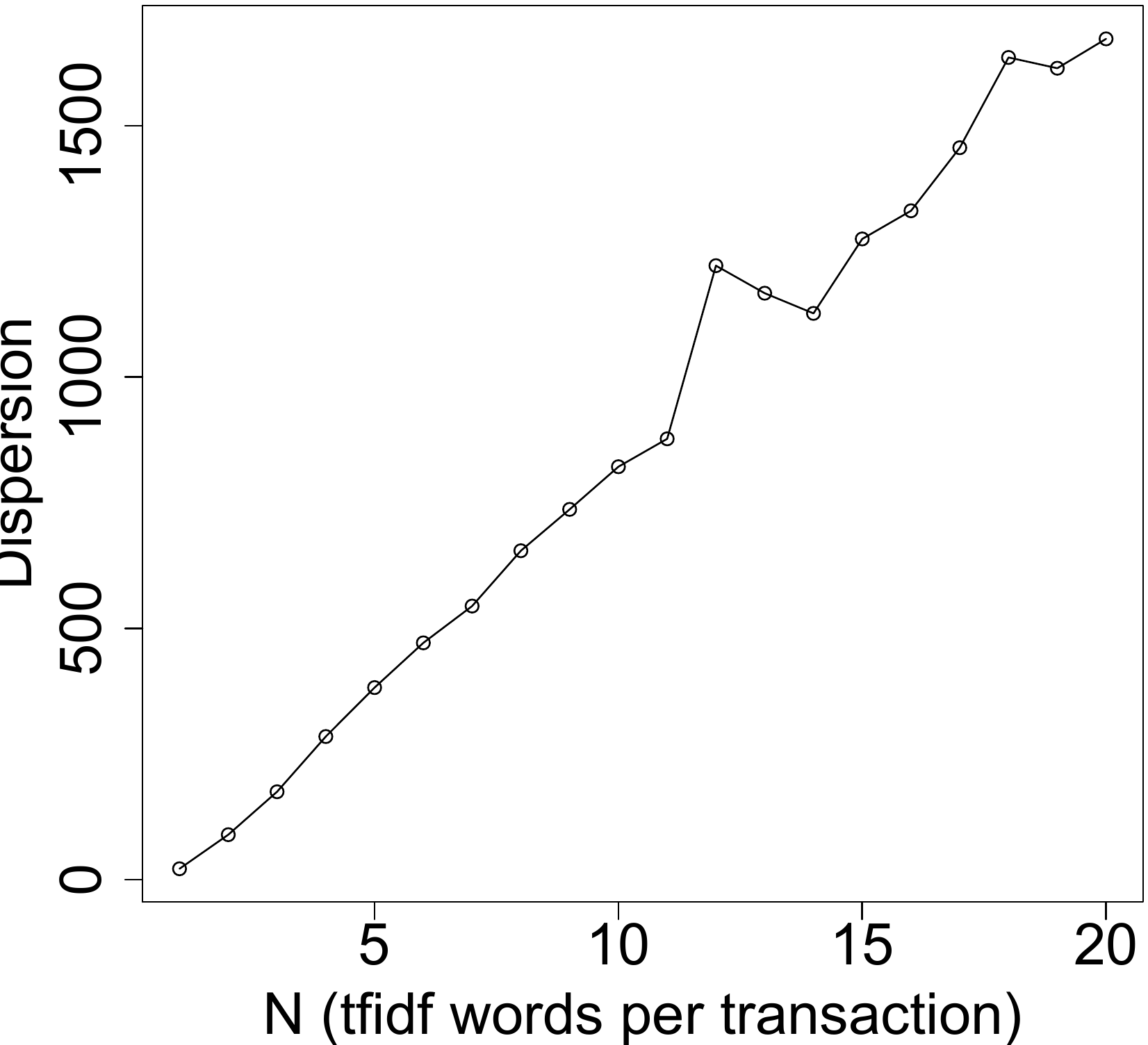}}
\caption{F-measure, variety and dispersion of tfidf-based pruning methods as a function of the number of words kept in the corpus\label{tfidf-fig}}
\end{figure}

Having calculated $\tfidf_S(w)$ for every $w\in S\in\mathbf{S}$, we take $N$ words from each sentence with the highest tfidf values, and use them as transaction items. The performance of this method depends on the value of $N$. On Fig.~\ref{tfidf-fig} the reader can see the values of three quantities as functions of $N$:
\begin{enumerate}
\item F-measure: we see that F-measure increases steadily and reaches a maximum value of 83.99 for $N=10$. Building transactions of more than 10 words (in decreasing tfidf order) deteriorates performance, in terms of F-measure;
\item variety: the number of predicted classes for sentences of the same document progressively increases but globally remains relatively low, around 3.1, except for $N=12$ and $N=13$ where it reaches $4.17$;
\item dispersion: it increases steadily, with again a small outlier for $N=12$, probably due to the higher variety obtained for that value of $N$.
\end{enumerate}

Furthermore, each investigated method  will generate transactions of various sizes. It is fair to compare them with tfidf-based methods with similar transactions sizes. Therefore we will use the results displayed in Fig.~\ref{tfidf-fig} to compare the performance of subsequent methods with the one of the tfidf-based method of similar transaction size. Table~\ref{tab:results_tdfif_based_pruning_N1} presents the results obtained by applying the tdfif-based pruning method, with a single word per transaction\enlargethispage{.5\baselineskip} ($N=1$).


\begin{table}[h!]
\caption{Tfidf-based pruning, keeping a single word per transaction}
\noindent\begin{tabularx}{\textwidth}{l|CCCCCCC|C}
&E12&GCRIM&GDIP&GJOB&GPOL&GSPO&GVIO&AVG\\\hline
Recall&69.30&44.48&55.44&45.75&52.54&82.90&67.98&59.77\\
Precision&70.09&77.81&71.25&79.76&71.62&80.78&73.35&74.95\\
F-measure&69.69&56.60&62.36&58.15&60.61&81.83&70.56&65.69\\\hline
\multicolumn{9}{l}{\footnotesize MinSupp=0.006, MinConf=67.6, Var.=1.36, Disp.=21.53, AvgTransSize=1.00}
\end{tabularx}
\label{tab:results_tdfif_based_pruning_N1}
\end{table}

\subsection{Methods based on dependencies}\label{dep-sec}

\begin{algorithm}[tb]
\SetKwFunction{classify}{Classify}
\SetKwFunction{evaluate}{Evaluate}
\SetKwFunction{prune}{Prune}
\SetKwFunction{hyper}{Hyperonymize}
\SetKwFunction{lemma}{Lemmatize}
\SetKwFunction{apriori}{Apriori}
\SetKwFunction{partition}{Partition}
\SetKwFunction{shuffle}{Shuffle}
\SetKwFunction{tfidf}{Tfidf}
\SetKwFunction{sort}{Sort}
\SetKw{KwTo}{:=}
\SetKwProg{Fn}{}{:}{end}

\KwData{An annotated corpus $\mathbf{S}$ and a morphosyntactic contraint $\phi\colon S\to\{\mathrm{true},\mathrm{false}\}$}
\KwResult{The pruned corpus $\mathbf{S}'$}
\Fn(){\prune{$\mathbf{S}$, $\phi$}}{
$\mathbf{S}'$ \KwTo $\emptyset$\;
\For{$S\in \mathbf{S}$}{
$S'$ \KwTo $\emptyset$\;
\For{$w\in S$}{
\If{$\phi(w)=\mathrm{true}$}{$S'$ \KwTo $S'\cup\{w\}$}
}
$\mathbf{S}'$ \KwTo $\mathbf{S}'\cup\{S'\}$\;
}
}
\caption{Dependency-based corpus pruning}\label{dep-algo}
\end{algorithm}

\begin{subtables}
\begin{table}[ht]
\caption{Strategy I$_0$: Pruning by keeping only heads of sentences}

\kern-5pt

\noindent\begin{tabularx}{\textwidth}{l|CCCCCCC|C}
&E12&GCRIM&GDIP&GJOB&GPOL&GSPO&GVIO&AVG\\\hline
Recall&38.54&57.46&26.88&17.43&31.06&88.49&51.90&44.54\\
Precision&49.13&66.18&49.61&65.73&39.07&42.18&60.31&53.17\\
F-measure&43.19&61.51&34.87&27.55&34.61&57.13&55.79&44.95\\\hline
\multicolumn{9}{l}{\footnotesize MinSupp=0.004, MinConf=36.6, Var.=2.14, Disp.=47.84, AvgTransSize=1.00}
\end{tabularx}
\label{tab:dependency_head_sentence}

\medskip

\caption{Strategy I$_1$: Pruning by keeping only nsubj $\to$ head dependencies}

\kern-5pt

\noindent\begin{tabularx}{\textwidth}{l|CCCCCCC|C}
&E12&GCRIM&GDIP&GJOB&GPOL&GSPO&GVIO&AVG\\\hline
Recall&70.55&60.44&66.97&58.27&63.22&78.76&69.92&66.88\\
Precision&73.05&80.84&72.87&86.98&71.71&85.29&76.29&78.15\\
F-measure&71.78&69.17&69.80&69.79&67.19&81.89&72.97&71.80\\\hline
\multicolumn{9}{l}{\footnotesize MinSupp=0.007, MinConf=60.4, Var.=1.43, Disp.=41.71, AvgTransSize=1.04}
\end{tabularx}
\label{tab:dependency_nominal_subject}
\end{table}
\begin{table}[t]
\caption{Strategy I$'_1$: Pruning by keeping only ccomp $\to$ head dependencies}

\kern-5pt

\noindent\begin{tabularx}{\textwidth}{l|CCCCCCC|C}
&E12&GCRIM&GDIP&GJOB&GPOL&GSPO&GVIO&AVG\\\hline
Recall&57.33&33.82&25.31&16.96&21.74&47.62&59.60&37.48\\
Precision&37.83&47.55&38.50&42.06&34.62&57.05&54.17&44.54\\
F-measure&45.59&39.53&30.54&24.17&26.71&51.91&56.75&39.31\\\hline
\multicolumn{9}{l}{\footnotesize MinSupp=0.008, MinConf=34.4, Var.=1.97, Disp.=19.59, AvgTransSize=1.15}
\end{tabularx}
\label{tab:dependency_clausal_complement}

\medskip

\caption{Strategy I$_2$: Pruning by keeping only nouns at distance~1 from head}

\noindent\begin{tabularx}{\textwidth}{l|CCCCCCC|C}
&E12&GCRIM&GDIP&GJOB&GPOL&GSPO&GVIO&AVG\\\hline
Recall&80.75&75.92&73.24&68.59&70.59&95.55&77.96&77.51\\
Precision&73.21&83.51&75.35&89.86&73.67&80.52&77.36&79.07\\
F-measure&76.80&79.53&74.28&77.80&72.09&87.39&77.66&77.94\\\hline
\multicolumn{9}{l}{\footnotesize MinSupp=0.016, MinConf=51.6, Var.=2.43, Disp.=244.82, AvgTransSize=2.70}
\end{tabularx}
\label{tab:dependency_pos_tags}

\medskip

\caption{Strategy I$'_2$: Pruning by keeping only verbs at distance~1 from head}

\kern-5pt

\noindent\begin{tabularx}{\textwidth}{l|CCCCCCC|C}
&E12&GCRIM&GDIP&GJOB&GPOL&GSPO&GVIO&AVG\\\hline
Recall&54.32&62.50&44.37&20.41&27.58&91.39&67.68&52.61\\
Precision&49.58&65.98&48.44&78.39&46.69&43.57&63.84&56.64\\
F-measure&51.84&64.19&46.32&32.39&34.68&59.01&65.70&50.59\\\hline
\multicolumn{9}{l}{\footnotesize MinSupp=0.019, MinConf=30, Var.=4.00, Disp.=175.41, AvgTransSize=2.01}
\end{tabularx}
\label{tab:dependency_verbs_distance1_head}
\end{table}
\end{subtables}

In this section we investigate several strategies using the dependency structure of sentences. Our general approach (cf. Alg.~\ref{dep-algo}) keeps only words of $S$ that fulfill a given morphosyntactic constraint $\phi$. The following strategies correspond to various definitions of $\phi$. 

\subsubsection{Strategy I$_0$}

Our first strategy will be to keep only the head of each sentence (which, incidentally, is a verb in 85.37\% of sentences of our corpus).
This corresponds to the constraint $\phi(w)\equiv(w=\mathrm{root}(S))$. Results are given on Table~\ref{tab:dependency_head_sentence}.


Although the recall of GSPO is quite high (a possible interpretation could be that sports use very specific verbs), F-measure is quite low when we compare it to the one of the tfidf-based method of the same average itemset length, namely~65.69\%.

\subsubsection{Strategy I$_1$} The second strategy consists in keeping words connected to the head by a (single) dependency of type nsubj (=~nominal subject). This
occurs in 79.84\% of sentences of our corpus. The constraint is then $\phi(w)\equiv(\exists(w,\mathrm{root}(S),$ $\mathrm{nsubj})\in \mathrm{Dep}(S))$. Results are given on Table~\ref{tab:dependency_nominal_subject}.




Note that the slightly higher than~1 transaction size is probably due to the rare cases where there are more than one nsubj dependencies pointing to the head. The scores rise dramatically when compared to those of the strategy based only on the head of the sentence. The average F-measure (71.80\%) is significantly higher than the tfidf-based performance for the same average transaction size (65.69\%). \emph{This shows that using a dependency property to select a word is a better choice than the one provided by the frequentist tfidf-based method.} Note that the case of nsubj is unique: if we take ccomp (=~clausal complement) instead of nsubj, the performance falls even below the level of strategy I$_0$ (Table~\ref{tab:dependency_clausal_complement}).



\subsubsection{Strategy I$_2$}
The third strategy considers all nouns (POS tags starting with \texttt{N}) at distance~1 from the head in the dependency graph. Such dependencies occur in 59.24\% of the sentences of our corpus. This corresponds to $\phi(x)\equiv((\exists(x,\mathrm{root}(S),d)\in \mathrm{Dep}(S))\wedge(\mathrm{POS}(x)=\mathtt{N*}))$. Results are given on Table~\ref{tab:dependency_pos_tags}.




The result seems better than the one of strategy I$_1$ (Table~\ref{tab:dependency_nominal_subject}). However, if we take transaction size into account, it is in fact merely equivalent to---and hence not better than, as it was the case for I$_1$---the tfidf-based method with the same transaction size. Once again we see a very high recall rate for the sports category.

One could be tempted to check the performance of taking verbs (instead of nouns) at distance~1 from the head. Indeed, verbs at that position are more frequent than nouns: they occur in 62.94\% of the sentences of our corpus. Nevertheless, the results are not as good (Table~\ref{tab:dependency_verbs_distance1_head}). This shows that despite their high frequency, verbs contain less pertinent information than nouns at the same distance from the head.









\subsection{Methods based on dependencies and hyperonyms}\label{hyper-sec}

\begin{algorithm}[tb]
\SetKwFunction{classify}{Classify}
\SetKwFunction{evaluate}{Evaluate}
\SetKwFunction{prune}{Prune}
\SetKwFunction{hyper}{Hyperonymize}
\SetKwFunction{lemma}{Lemmatize}
\SetKwFunction{apriori}{Apriori}
\SetKwFunction{partition}{Partition}
\SetKwFunction{shuffle}{Shuffle}
\SetKwFunction{tfidf}{Tfidf}
\SetKwFunction{sort}{Sort}
\SetKw{KwTo}{:=}
\SetKwProg{Fn}{}{:}{end}

\KwData{A dependency-pruned corpus $\mathbf{S}'$, an hyperonymic function $\mathrm{MSCH}\colon \mW\to \mW^{\mathbb{N}}$, the hyperonymic order $N$}
\KwResult{The hyperonymically extended corpus $\mathbf{S}''$}
\Fn(){\hyper{$\mathbf{S}'$, $\mathrm{MSCH}$, $N$}}{
$\mathbf{S}''$ \KwTo $\emptyset$\;
\For{$S'\in \mathbf{S}'$}{
$S''$ \KwTo $\emptyset$\;
\For{$w\in S'$}{
\eIf{$\mathrm{proj}_N(\mathrm{MSCH}(w))\ne\emptyset$}{$S''$ \KwTo $S''\cup\{\mathrm{proj}_N(\mathrm{MSCH}(w))\}$}{$S''$ \KwTo $S''\cup\{w\}$}
}
$\mathbf{S}''$ \KwTo $\mathbf{S}''\cup\{S''\}$
}
}
\caption{Corpus hyperonymization}\label{hyper-algo}
\end{algorithm}

In this section we add semantic information by the means of hyperonyms, using the hyperonymization function $h$ (\S\,\ref{hc}).
The preprocessing is done by Alg.~\ref{hyper-algo}: $h_i(w)$ is an $N$-th order hyperonym of $w$, if it exists in WordNet. In case there is no $N$-th order hyperonym, the word remains unchanged. We call $N$ the \emph{hyperonymic factor} of our itemset transformation.

\subsubsection{Strategy II$_1$} This strategy considers hyperonymic factor $N=1$. We thus first apply strategy I$_1$ and then hyperonymization $h_1$. Results are presented on Table~\ref{tab:hyperonymic_factor_N1}.


\begin{subtables}
\begin{table}[t]
\caption{Strategy II$_1$: I$_1$ followed by first-order hyperonymization}

\kern-5pt

\noindent\begin{tabularx}{\textwidth}{l|CCCCCCC|C}
&E12&GCRIM&GDIP&GJOB&GPOL&GSPO&GVIO&AVG\\\hline
Recall&72.39&56.04&71.32&48.96&59.02&82.20&70.42&65.76\\
Precision&66.21&75.42&64.33&82.21&67.71&73.20&70.73&71.40\\
F-measure&69.16&64.30&67.64&61.37&63.07&77.44&70.57&67.65\\\hline
\multicolumn{9}{l}{\footnotesize MinSupp=0.010, MinConf=44.8, Var.=1.92, Disp.=78.47, AvgTransSize=1.04}
\end{tabularx}
\label{tab:hyperonymic_factor_N1}

\medskip

\caption{Strategy II$_2$: I$_1$ followed by second-order hyperonymization}

\kern-5pt

\noindent\begin{tabularx}{\textwidth}{l|CCCCCCC|C}
&E12&GCRIM&GDIP&GJOB&GPOL&GSPO&GVIO&AVG\\\hline
Recall&69.02&52.78&71.97&47.77&54.24&80.80&65.67&63.18\\
Precision&64.94&74.57&61.23&80.64&65.81&69.96&72.33&69.93\\
F-measure&66.92&61.81&66.16&60.00&59.47&74.99&68.84&65.46\\\hline
\multicolumn{9}{l}{\footnotesize MinSupp=0.008, MinConf=44.8, Var.=1.85, Disp.=70.51, AvgTransSize=1.04}
\end{tabularx}
\label{tab:hyperonymic_factor_N2}
\end{table}
\end{subtables}

The performance is globally inferior to the one of Strategy I$_1$ (in which, F-measure attained 71.80\%). It is interesting to note that the recall of class GJOB has decreased significantly (48.96\% vs. 58.27\%): in other words, using hyperonyms when dealing with labor issues results into failure to recognize 9.31\% of the documents as belonging to the domain;  one could say that terms used in GJOB lose their “labor specificity” already at first-order hyperonymization. On the other hand, the (already high in I$_1$) recall of GSPO has increased even more, compared to I$_1$ (from 78.76\% to 82.20\%): it seems that sports terminology remains in the domain even after hyperonymization, and replacing specific terms by more general ones has increased their frequency as items, and hence improved recall. We have the same phenomenon with the recall of GDIP (which increased from 66.97\% to 71.32\%), and also slightly with the recalls of E12 and GVIO.

\subsubsection{Strategy II$_2$} This strategy is similar to strategy II$_1$ but uses hyperonymic factor $N=2$. Results are presented on Table~\ref{tab:hyperonymic_factor_N2}.



The performance is globally inferior to the one of II$_1$ (where we used first-order hyperonyms), with two minor exceptions: the recall of GDIP that increased by 0.65\% and the precision of GVIO that increased by 1.6\%. What is noteworthy however, is the fact that the recalls of GDIP and GSPO are still higher than the ones of strategy I$_1$ (no hyperonyms).

\begin{figure}[ht]
\centering
\resizebox{\textwidth}{!}{\includegraphics{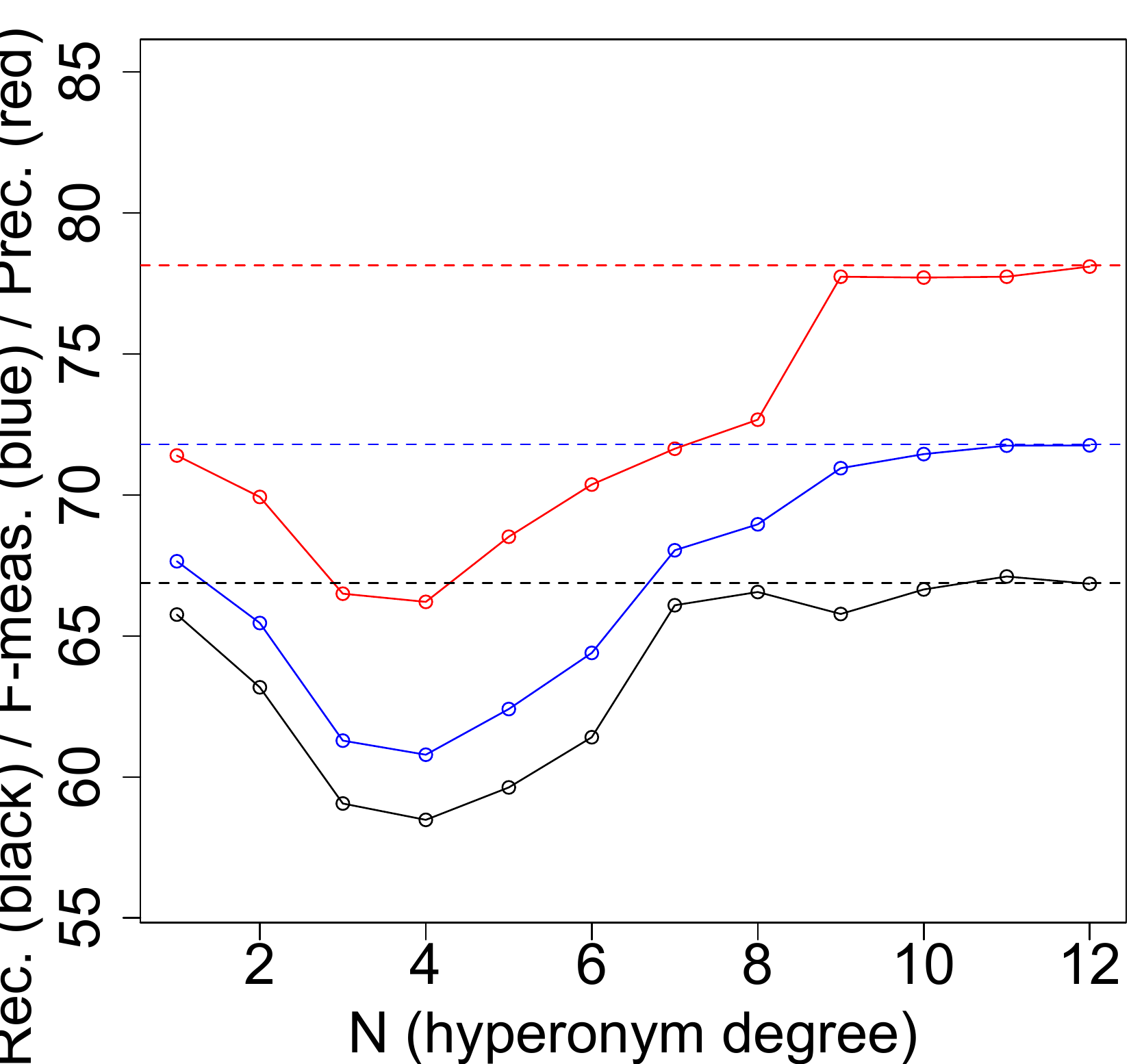}~\includegraphics{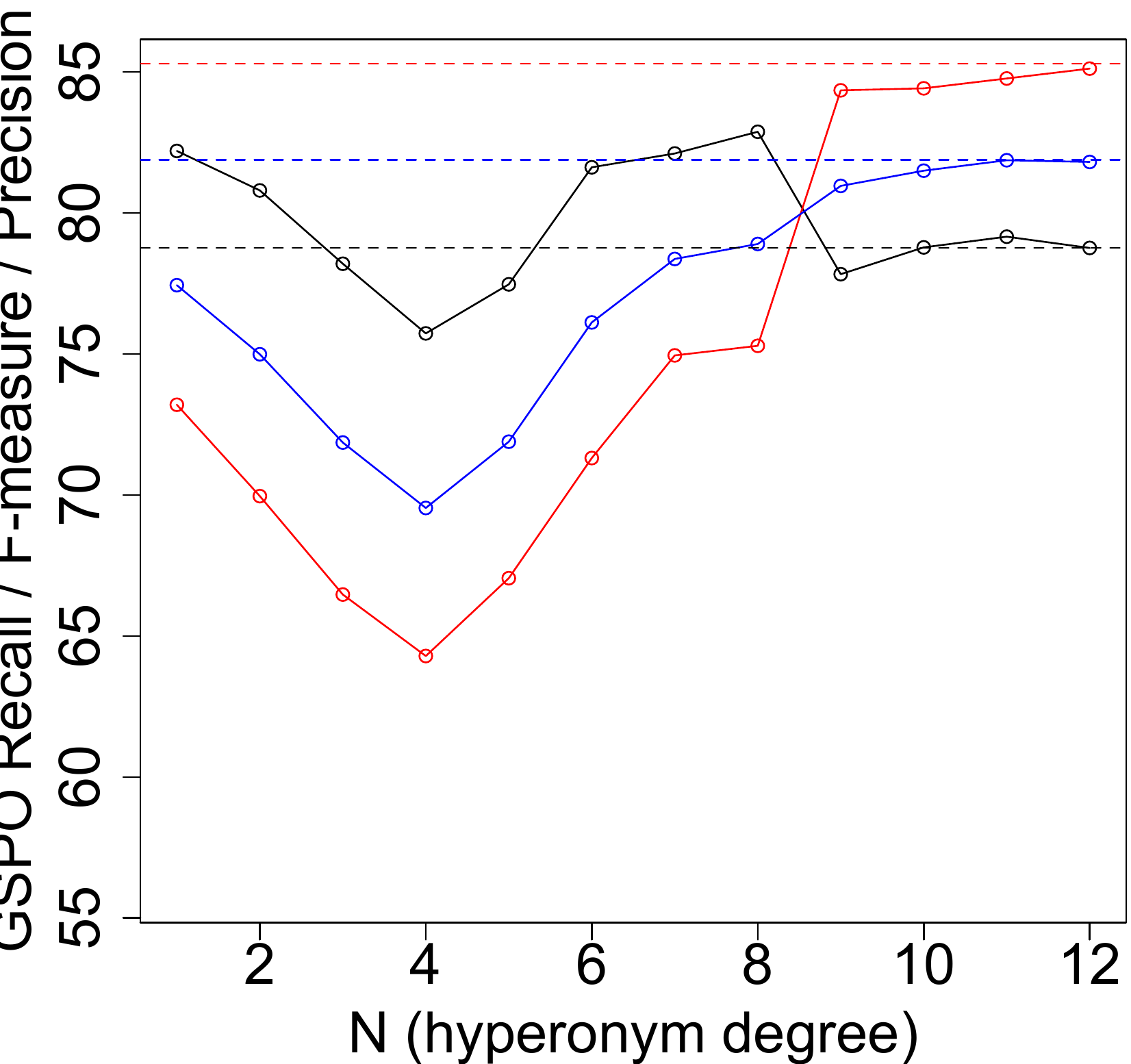}~\includegraphics{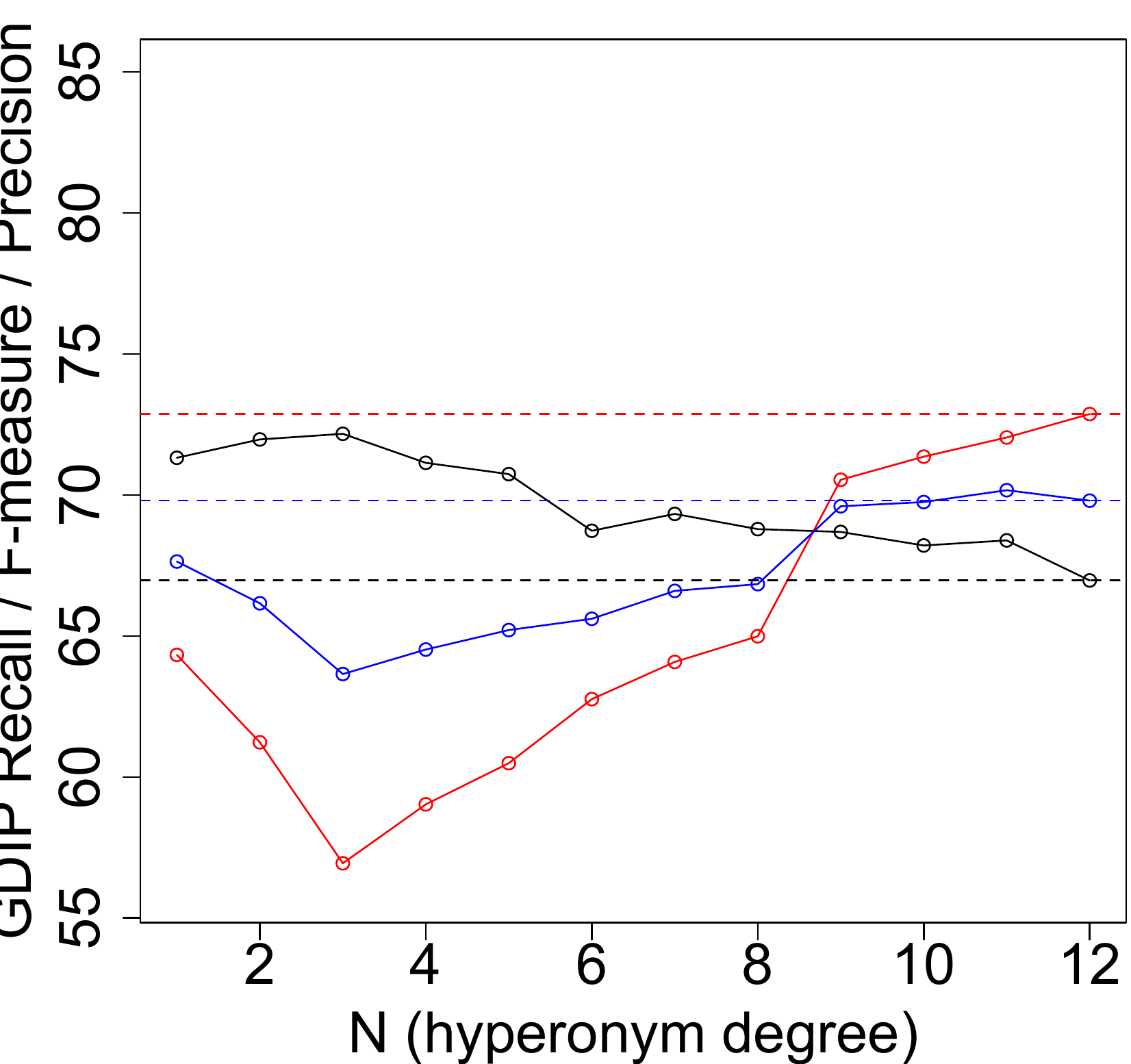}}
\caption{F-1 measure for hyperonymization of orders $1\leq N\leq12$: the average case, class GSPO, class GDIP\label{hypers-fig}}
\end{figure}

To better understand the behavior of the system when climbing the hyperonymic chain by replacing words by hyperonyms of increasingly higher order (and returning to the original word when there are no hyperonyms left) we calculated the performance for $N$-th order hyperonyms for $1\leq N\leq12$. Note that when $N>12$ the amount of remaining hyperonyms is negligible and the strategy is similar to strategy I$_1$ (no hyperonyms). On Fig.~\ref{hypers-fig}, the reader can see the evolution of recall (black), precision (red) and F-measure (blue) for the average of all class, and then specifically for GSPO and for GDIP. Dashed lines represent the recall, precision and F-measure of strategy I$_1$.

In the average case, the effect of hyperonymization of orders 1--4 is to decrease performance. After $N=5$, the global number of hyperonyms available in WordNet rapidly decreases so that the situation gradually returns to the one of I$_1$ (no hyperonyms) and we see curves asymptotically converging to I$_1$ lines from underneath. 

Not so for GSPO, the GSPO recall curve of which is above the I$_1$ value for most  $N$ ($N=1$, 2, 6--8 and 10--12). 

The phenomenon is even better illustrated in the case of GDIP: as the reader can see on the figure, the \emph{complete GDIP recall curve is located above the I$_1$ one}.
It seems that in these two cases (GDIP and, to a lesser extent, GSPO), hyperonyms of all orders have a positive impact on the classifier. Unfortunately this impact only concerns recall and is compensated by bad precision, so that F-measure is still inferior to the I$_1$ case.

\section{Conclusion and future work}\label{con}

In this paper we have investigated the use of association rules for text classification by applying two new techniques: (a) we reduce the number of word features through the use of morphosyntactic criteria in the framework of dependency syntax; for that we keep words dependent from the head by specific dependencies and/or having specific POS tags (b) we replace words by their hyperonyms of different orders, which we have calculated out of WordNet using frequencies and, in some cases, disambiguation. We have obtained positive results for case (a), in particular when we compare dependency-based single-item rules with tfidf-based ones. In case (b) the results we share in this paper are less efficient but still interesting, especially we found classes for which hyperonymization significantly improves recall.

This work opens several perspectives, among which: 

--- examine why these particular classes are favorable to hyperonymization, whether this is related to the structure of WordNet or to linguistic properties of the domain;

--- explore \emph{partial hyperonymization} i.e., is it possible to hyperonymize only specific items according to the needs of the classifier?\footnote{Indeed, by hyperonymizing all words one wins on one side and loses on the other: for example “dalmatian” and “poodle” will both be replaced by “dog”, but “dog” occurrences will be replaced by “canid”. It would be more preferable to keep the word “dog” in the second case, so that we have a real increase in frequency.} How do we choose, on the word level, if we should rather keep the original word (to increase precision) or switch to some hyperonym (to increase recall)?



--- we have used only recall and precision as quality measures of our rules, and our evaluation is strongly dependent on these measures since the selection of the 1,000 rules we keep is entirely based upon them. There are other quality measures available, how do they apply and how can they be compared and combined? How robust are the results? 

--- and finally: how can we optimize the distinctive feature of association rules, namely the fact of being intelligible by the user? How can the user's experience (and linguistic knowledge) be incorporated in the enhancement of rules to obtain the best possible result from his/her point of view?


\bibliographystyle{splncs03}
\bibliography{main}

\begin{thebibliography}{10}
\providecommand{\url}[1]{\texttt{#1}}
\providecommand{\urlprefix}{URL }

\bibitem{reuters}
Reuters corpus, volume~1, english language, 1996-08-20 to 1997-08-19,
  \url{http://about.reuters.com/researchandstandards/corpus/statistics/index.asp}

\bibitem{bnc}
British {N}ational {C}orpus (1994), \url{http://www.natcorp.ox.ac.uk}

\bibitem{aggarwal_zhai_MTD_2012}
Aggarwal, C.C., Zhai, C.: A Survey of Text Classification Algorithms, chap.~6,
  pp. 163--222. Mining Text Data, Springer

\bibitem{ahonen_etal_TR_1997}
Ahonen, H., Heinonen, O., Klemettinen, M., Verkamo, A.I.: Applying data mining
  techniques in text analysis. TR C-1997-23, Department of Computer Science,
  University of Helsinki (1997)

\bibitem{apriori}
Borgelt, C.: Efficient implementations of apriori and eclat. In: {Workshop of
  Frequent Item Set Mining Implementations (FIMI 2003), Melbourne, FL, USA}
  (2003)

\bibitem{FerreriCancho:2004wd}
Ferrer~i Cancho, R., Sol{\'e}, R.V., K{\"o}hler, R.: {Patterns in syntactic
  dependency networks}. Physical Review E  69,  1--8 (2004)

\bibitem{cherfi_etal_PMAR_2009}
Cherfi, H., Napoli, A., Toussaint, Y.: A Conformity Measure Using Background
  Knowledge for Association Rules: Application to Text Mining, pp. 100--115.
  IGI Global (2009)

\bibitem{chomsky1957}
Chomsky, N.: Syntactic structures. Mouton (1957)

\bibitem{cohen_AAAI_1996}
Cohen, W.W.: Learning rules that classify e-mail. In: AAAI Spring Symposium on
  ML and IR (1996)

\bibitem{Curran:2002vm}
Curran, J.R., Moens, M.: {Scaling context space}. In: Proceedings of the 40th
  Annual Meeting of the Association for Computational Linguistics. pp. 231--238
  (2002)

\bibitem{do_Master_2012}
Do, T.N.Q.: {A graph model for text analysis and text mining. Master Thesis,
  Universit{\'e} de Lorraine, (2012)}

\bibitem{jaillet_etal_IDA_2006}
Jaillet, S., Laurent, A., Teisseire, M.: Sequential patterns for text
  categorization. Intell. Data Anal.  10(3),  199--214 (2006)

\bibitem{Kovacs:2008da}
Kovacs, L., Baksa-Varga, E.: {Dependency-based mapping between symbolic
  language and Extended Conceptual Graph}. In: 6th International Symposium on
  Intelligent Systems and Informatics. pp. 1--6 (2008)

\bibitem{lang_ICML_1995}
Lang, K.: Newsweeder: Learning to filter netnews. In: International Conference
  on Machine Learning. pp. 331--339. Morgan Kaufmann (1995)

\bibitem{CAR}
Liu, B., Hsu, W., Ma, Y.: Integrating classification and association rule
  mining. In: Proc. of the {Int. Conf. on Knowledge Discovery and Data Mining}
  ({New York}). pp. 80--86 (1998)

\bibitem{Lowe:2001wx}
Lowe, W.: {Towards a theory of semantic space}. In: Proceedings of the
  Twenty-Third Annual Conference of the Cognitive Science Society. pp. 576--581
  (2001)

\bibitem{sdp}
de~Marneffe, M.C., MacCartney, B., Manning, C.D.: Generating typed dependency
  parses from phrase structure parses. In: LREC 2006. pp. 449--454 (2006)

\bibitem{shami_etal_AAAI_1998}
Mehran~Sahami, Susan~Dumais, D.H., Horvitz, E.: A bayesian approach to
  filtering junk email. In: AAAI Workshop on Learning for Text Categorization.
  AAAI Technical Report WS-98-05 (1998)

\bibitem{melcuk}
Mel$'$\v{c}uk, I.A.: Dependency syntax : theory and practice. Albany: State
  University Press of New York (1987)

\bibitem{miller1995}
Miller, G.A.: {WordNet}: {A} lexical database for {E}nglish. Communications of
  the ACM  38(11),  39--41 (1995)

\bibitem{nivre_TR_2005}
Nivre, J.: {Dependency Grammar and Dependency Parsing. MSI report 05133. School
  of Mathematics and Systems Engineering, V{\"a}xj{\"o} University, (2005)}

\bibitem{OrdonezSalinas:2010db}
Ordo{\~n}ez-Salinas, S., Gelbukh, A.: {Information Retrieval with a Simplified
  Conceptual Graph-Like Representation}  6437(Chapter 9),  92--104 (2010)

\bibitem{Pado:2007bu}
Pad{\'o}, S., Lapata, M.: {Dependency-based construction of semantic space
  models}. Computational Linguistics  33(2),  161--199 (2007)

\bibitem{pedersen}
Pedersen, T.: Information content measures of semantic similarity perform
  better without sense-tagged text. In: {Proceedings of the 11th Annual
  Conference of the North American Chapter of the Association for Computational
  Linguistics (NAACL HLT 2010)}. pp. 329--332 (2010)

\bibitem{roche_etal_IIPWM_2004}
Roche, M., Az{\'e}, J., Matte-Tailliez, O., Kodratoff, Y.: Mining texts by
  association rules discovery in a technical corpus. In: Intelligent
  Information Processing and Web Mining. pp. 89--98. Advances in Soft
  Computing, Springer Verlag (2004)

\bibitem{treetagger}
Schmid, H.: Probabilistic part-of-speech tagging using decision trees. In:
  {Proceedings of International Conference on New Methods in Language
  Processing, Manchester, UK} (1994)

\bibitem{sebastiani_ACM-CS_2002}
Sebastiani, F.: Machine learning in automated text categorization. ACM Comput.
  Surv.  34(1),  1--47 (2002)

\bibitem{tesniere1959}
Tesni\`ere, L.: {\'E}l\'ements de syntaxe structurale. Klincksieck (1959)

\bibitem{zaiane_antonie_ADC_2002}
Za\"{\i}ane, O.R., Antonie, M.L.: Classifying text documents by associating
  terms with text categories. In: Australasian Database Conference. CRPIT,
  vol.~5. Australian Computer Society (2002)

\end{thebibliography}

\end{document}